\documentclass[preprint,12pt]{elsarticle}

\usepackage{amssymb}

\usepackage{setspace}

\begin{document}

\begin{frontmatter}



\title{Understanding bond formation and its impact on the capacitive properties of SiW12 polyoxometalates adsorbed on functionalized Carbon Nanotubes}


\author[IER]{Alfredo Guill\'en-L\'opez}
\author[IER]{N\'estor Espinosa-Torres}
\author[IER]{Ana Karina Cuentas-Gallegos}
\author[IER]{Miguel Robles}
\author[IER,CONACYT]{Jes\'us Mu\~{n}iz\footnote{Corresponding author. E-mail: jms@ier.unam.mx (Jes\'us Mu\~{n}iz). Tel.: +52 (777) 362 0090}}

\address[IER]{Instituto de Energ\'ias Renovables, Universidad Nacional Aut\'onoma de M\'exico, Priv. Xochicalco s/n, 
              Col.  Centro, Temixco, Morelos. CP 62580, Mexico}
              
\address[CONACYT]{CONACYT-Universidad Nacional Aut\'onoma de M\'exico, Priv. Xochicalco s/n, Col.  Centro, Temixco, Morelos. CP 62580, Mexico}

\begin{abstract}

In the last decade, a large number of studies at the experimental level in electrochemical systems for energy storage devices have been performed. However, theoretical approaches are highly desirable to understand the physicochemical properties giving rise to energy storage phenomena. This work was intended to provide insights into the $in$ $silico$ design of novel nanocomposite materials formed by the Keggin polyoxometalate SiW12 anchored to an organic functional group $\varphi-X$ (with $X= -NH_2, -OH, -COH$ and $-COOH$) linked to a carbon nanotube. In these systems, the Density of States around the Fermi level is enhanced, giving the composite material the capacity of facile electron transport that may be determinant at the charge/discharge cycling performed in energy storage devices. Charge transfer at the composite materials under study is greatest for the $\varphi-COOH$ functional group, yielding an attraction with the SiW12 cluster of the same order of magnitude as that of covalent nature. The rest of the functional groups induce a non-covalent interaction of the electrostatic-type, mediated by a van der Waals attraction. Our proposed methodology may represent a tool to develop novel electrode materials that may improve the performance on energy storage devices, such as supercapacitors or Li-ion batteries.

\end{abstract}


\end{frontmatter}

\section{Introduction}
\label{Introd}

The chemistry of polyoxometalates (POM) comprises metal oxides of the basic form $[MO_x]_n$, where M = Mo, W, V, Nb, and $x$ = 4-7 \cite{Long:10}. A large number of combinations for POM formation may be achieved within this framework. Particularly, the class of heteropolyanions present 2 different atoms, namely  the heteroatom and the addenda atom. This classification includes the Keggin-symmetry $[XM_{12}O_{40}]^{n-}$ (or simply XM12), and the Dawson-symmetry $[X_2M_{18}O_{62}]^{n-}$, where X represents the heteroatom and M represents the addenda atom. The Keggin POMs present potential properties to be implemented in energy storage applications and also in renewable energy materials\cite{Luo:12}. Such clusters are highly stable during their fast multielectron oxidation/reduction processes and may play an important role in electron transfer mechanisms\cite{Sadakane:98}. Additionally, the POM properties may be designed since the redox activity is sensitive to the substitution of the hetero or addenda atoms. This POM capability, mixed with the ability to transfer electronic charge, as well as its facile adsorption on organic/inorganic surfaces, may be determinant to tailor novel electrodes for energy storage devices. The development of electrodes based on POM$@$Carbon substrates for energy storage devices, includes several techniques such as immobilization in a polymer matrix, layer-by-layer self-assembly, and chemisorption on carbon substrates. This last technique involves an energetic stable and irreversible bonding\cite{Kang:04,Garrigue:04,Kulesza:06}, which suggests that the design of several POM$@$Carbon materials may be achieved. The route of preparation involves the oxidation of the carbon substrate by using an acid to provide functional groups that act as active sites where the POM may interact. The carbon substrate is further dispersed in the POM solution to promote its immobilization. This methodology may be applied on different carbon substrates such as carbon nanofibers (CNF)\cite{Cuentas:07a}, activated carbon (AC) \cite{Ruiz:12,Suarez:14,Cuentas:16}, graphene \cite{Tessonnier:13,Kume:14,Li:10,Zhou:10} or Multi-walled carbon nanotubes (MWCNT) \cite{Kang:04,Skunik:08,Cuentas:09,Cuentas:06,Kawasaki:11,Li:11,Fei:06,Cuentas:07}. Our experimental group has reported the synthesis of POM$@$MWCNT \cite{Cuentas:06}, POM$@$CNF \cite{Cuentas:07a}, and more recently, POM$@$AC\cite{Cuentas:16}. The experimental analysis showed \cite{Cuentas:06,Cuentas:07a} that the acid treatments induced the formation of oxygen functional groups at the surface, such as hydroxyl, carbonyl and carboxylic acid. Based on FTIR spectra, higher POM concentration was achieved for nanocarbons treated with a stronger acidic procedure, which showed higher carbonyl group concentration that could be related with ketone, aldehyde or carboxylic acid groups. The resulting composite material presented the most homogeneous distribution of the PMo12 clusters, with the largest dispersion and highest loading \cite{Cuentas:06}. This suggests that the carboxylic groups may provide to the carbon substrate, available locations where the POM may be tightly anchored, such as it has extensively been studied with FTIR analysis\cite{Cuentas:07a,Cuentas:06,Kang:04,Skunik:08}. Consequently, the mechanism behind the chemisorption of POM$@$Carbon substrates, was suggested to rise from electron and POM reduction at the $-OH$ fragment.

On the other hand, the absence of carbon oxidation may lead to unsuccessful POM retention, as it was previously performed\cite{Kang:04} in the composite material $[P_2W_{18}O_{62}]^{6-}/CNT$, where no POM adsorption was reported due to the lack of carbon oxidation. The anchoring of POMs on carbon has also been applied on ACs, such as the chemisorption of PMo12 and $[PW_{12}O_{40}]^{3-}$ on Norit DLC Super 30 activated carbon \cite{Ruiz:12,Suarez:14}, where a 50\% of increasing in mass was observed after POM adsorption. Furthermore, Cuentas-Gallegos et al\cite{Cuentas:16} incorporated PMo12 on Vulcan carbon functionalized with $\varphi-OH$ and $\varphi-NH_2$. A strong chemisorption was evidenced. This was specifically observed for the $\varphi-OH$ groups. Furthermore, the $PMo12@\varphi-OH/VC$ system showed the best performance with enhanced density currents in the cyclic voltammograms. The POM anchoring has also been extended to obtain POM$@$Graphene systems by sonicating highly ordered pyrolitic graphite (HOPG) in an aqueous solution of PMo12 \cite{Fattakhova:06}. The strong interaction of POM with the first graphene sheet, widens the interplanar distance of graphite, which allows to exfoliate the rest of the graphene layers. Cuentas-Gallegos et al \cite{Cuentas:11} synthesized the SiW12$@$MWCNT system. FTIR probes certified the anchoring of SiW12 clusters on the MWCNTs. The microstructure and dispersion were correlated with BET surface area. Consequently, it was evidenced that the SiW12 clusters are firmly adsorbed to the carbon substrate. The synthesis of POM$@$MWCNT was also reported \cite{Cuentas:10}. It was shown that higher dispersion improves electrochemical properties. Nevertheless, it was evidenced that redox processes at PMo12$@$MWCNT \cite{Cuentas:10} system are scarce, on the contrary to the SiW12$@$MWCNT \cite{Cuentas:11} case, where they were abundant. Genovese and Lian \cite{Genovese:16} reported the composite GeMo12$@$CNT material, fabricated with PDDA (polyelectrolyte linker poly-diallyldimethylammonium chloride). This gives a low POM loading with high conductivity and improved performance at charge/discharge cycles. The same material was also synthesized with a PIL (Polymerized Imidazolium Linker) based on monomer of ethyl-vinyl-imidazolium, yielding a higher POM loading and acceptable conductivity. This suggests that the inclusion of discrete organic functional groups into the POM/CNT interface may increase desirable properties on energy storage devices.

In the theoretical field, only a few studies are available in open literature regarding the nature of bonding in this kind of composite materials. For instance, Wen et al \cite{Wen:12} studied the electronic structure properties of the PW12$@$Graphene system by using Density Functional Theory (DFT) at the GGA(General Gradient Approach) and LDA(Local Density Approach) levels. The adsorption energies of the PW12 anchored to graphene were studied, and also the charge transfer properties. A charge transfer coming from PW12 to graphene was predicted. The same group\cite{Wen:13} also studied the PMo12$@$SWCNT system by using DFT at GGA/PBE level of theory. Adsorption energies and charge transfer mechanisms were also analyzed by considering SWCNT of semiconducting and metallic nature. Such charge transfer is highly sensitive to the nature of the SWCNT. Mu\~{n}iz et al\cite{Muniz:16} studied the composite systems PMo12$@$Graphene and $PMo12@\varphi-X/Graphene$, with $X=-OH$ and $-NH_2$. It was found that the POM cluster interacts with the carbon substrate through non-covalent attractions of the electrostatic type. The functionalized $PMo12@\varphi-OH/Graphene$ system was found to be the most capable to retain electronic charge, which may be readily ascribed to an enhanced availability to yield higher density currents in cyclic voltammogram studies. This study was extended on the series of systems $[XM_{12}O_{40}]^{n-}@Graphene$ \cite{Muniz:17}, where $X=Pd, P, Ru, Si$; $M=Mo,Nb,W$ and $n=2,3,4,8$, respectively. In this comparative DFT study, different anchoring mechanisms of POM at the graphene substrate were revised. The prediction of the POMs PdMo12 and RuNb12 was anticipated, revealing that at the composite carbon interface, enhanced energy storage properties may be present, due to the increased number of Density of States around the Fermi level; which may promote the transit of ions in the charge/discharge processes observed in energy storage devices. Particularly, the SiW12$@$Graphene systems show a charge transfer mechanism where the electronic charge migrates from the SiW12 cluster to the graphene layer\cite{Muniz:17}. Recently, Lang et al\cite{Lang:17} performed a DFT study on the lacunary species $[PW_{11}O_{39}]^{7-}$ adsorbed on Au(100) and Ag(100); that is, PW11$@$Au(100) and  PW11$@$Ag(100). It was found that the incorporation of counterions is critical to reproduce all properties of the composite, adequately. It was also found that the PW11 cluster is adsorbed through the external O-atoms in a similar orientation as that found for the $[SiW_{12}O_{40}]^{4-}$ system. Aparicio-Angl\`{e}s et al \cite{Aparicio:12} performed a theoretical study combining classical Molecular Dynamics (MD) and DFT on the SiW12@Ag(100) system. The inclusion of water solvent molecules in the simulation and the presence of counterions reveal a charge transfer from the Ag surface to the SiW12 cluster, evidencing its immediate reduction. The SiW12 adsorption on the metal was also studied by plane waves at the DFT level and the IR spectrum was predicted \cite{Aparicio:11}, which allowed to characterize the signal associated to the stretching mode W-O-Ag located around 800 cm$^{-1}$. Rozanska et al\cite{Rozanska:11} studied at DFT level the $PMo12@\beta-crystobalite$ (001), (101) and (111). A covalent bonding was found between both systems and a weaker covalent bonding is formed if the surface is partially dehydroxylated, which plays a crucial role on the properties of anchored POMs. The study of composite systems where metal oxides are grafted on carbon substrates has been widely extended to POM$@$Carbon nanocomposite materials, but fundamental studies at the atomistic level, where the mechanism behind the bonding at the interface metal oxide/carbon substrate is evidenced, are scarce. The study of such systems is only limited to a small number, and the profound understanding of the bonding properties and also the electronic structure properties that may relate to energy storage capabilities is of relevant concern. The aim of this study is to give insights into the nature of bonding at the interface SiW12 cluster/CNT, via the grafting of organic functional groups, and also to elucidate and predict those functional groups that may be potential candidates to be implemented at the SiW12/CNT interface that may maximize the related-energy storage capabilities.

\section{Computational Details}
\label{Comp}

All systems under study were fully optimized by considering periodic boundary conditions and using Density Functional Theory at the Generalized Gradient Approximation (GGA) with the PBE (Perdew-Burke-Ernzerhof) functional (DFT/PBE). The basis set used was the Troullier - Martins non-local form of norm-conserving pseudopotentials and localizad atomic orbitals. The double - $\zeta$ plus polarization basis set (DZP) was considered in all elements of the systems under study. We used all pseudopotentials as given in the SIESTA $ab$ $initio$ package \cite{Ordejon:96,Portal:97,Soler:02}. On the other hand, the influence of dispersive interactions coming from the van der Waals (vdW) attractions may also be involved at a significant extent, since it has been shown that its inclusion at organic/inorganic systems\cite{Liu:14} may be important. We used the PBE + vdW functional\cite{Tkatchenko:09,Carrasco:14}, developed by Tkatchenko et al\cite{Tkatchenko:09}, whose methodology is based upon a pair-wise atom-atom approach. In the calculations performed within this framework, numeric atom-centered basis sets were used, as implemented in the all-electron code FHI-aims\cite{Carrasco:14,Blum:09,Havu:09}. In this computation, the introduction of the atomic Zeroth-Order Regular Approach (ZORA) was considered to include relativistic effects\cite{Lenthe:99} that may be relevant in the chemistry of POMs\cite{Lopez:12}. In this case, tight conditions were set, and the '$tier2$' standard basis set for C, O, N and O atoms were used in the computations, while the '$tier1$' basis set was used for the Si atom.

In all computations performed at DFT/PBE and DFT/PBE+vdW levels, a threshold of 0.01 eV{\AA}$^{-1}$ was imposed in the convergence criteria for the final forces at all geometrical optimizations, while 10$^{-5}$ electrons were implemented for the electron density, and 10$^{-5}$ eV for the total energy of the systems. With regard to the Mokhorst and Pack\cite{Monkhorst:76} grid, a 3 x 3 x 1 $k$ - point was introduced for the primitive cell to sample the Brillouin zone. The systems under study are based upon an extended (7,7) CNT with armchair symmetry, formed by 224 atoms. This configuration was chosen, since the size of this substrate is sufficient to model the interaction with another system, such as POM. Periodic boundary conditions were imposed in order to simulate a network that may contain the conditions of a fragment in an electrode of an energy storage device. All calculations were performed with a unit cell of dimension 40 x 25 x 70 {\AA}, allowing the composite systems to be restrained to a periodic lattice. The cell size was considered with a vacuum separation in the $z$-direction, which avoids spurious interactions with images in the periodic lattice.

The energetic stability was assessed with the calculations of adsorption energies ($\Delta E_a$), as given by Equation \ref{Eads}:

\begin{equation}
\Delta E_{a} = E_{T}-E_{\varphi-X/CNT}-E_{SiW12},
\label{Eads}
\end{equation}

where $E_T$ corresponds to the total energy of the different composite systems under study, including those where a pristine CNT is considered and those CNTs whose surfaces are functionalized by $X$ that corresponds to: $i)$ $\varphi-NH_2$, $ii)$ $\varphi-OH$, $iii)$ $\varphi-COH$, $iv)$ $\varphi-COOH$. The values $E_{\varphi-X/CNT}$ represent: the total energy of the pristine, isolated CNT in a unit cell, and the total energy of a functionalized CNT with one of the 4 functional groups cited above. $E_{SiW12}$ is the total energy of the isolated SiW12 cluster inside the same unit cell. Furthermore, the adsorption energies were also benchmarked with the GGA/PBE functional, using a van der Waals density functional, namely VDW-DF, as given by Dion et al\cite{Dion:04,Cooper:10}. This was performed to quantify the degree of vdW attraction present in the composite systems.

Charge transfer analysis among the functionalized CNTs and the SiW12 cluster was performed using Mulliken analysis, which it is well known to be highly sensitive to the basis set. Consequently, we also performed Hirshfeld and Voronoi schemes to verify Mulliken reliability. We implemented the method suggested by Arellano et al\cite{Arellano:00}  that considers the partition of the charge density in real space to determine a possible interaction and to evaluate a charge transfer between both SiW12 cluster and $\varphi-X/CNT$, where $X$ may correspond to the 4 cases given above, or it may represent no functional groups (pristine CNT). That is, we considered the charge transfer $\Delta Q$ as the difference of the Mulliken charges located at the isolated POM and those charges redistributed at the adsorbed POM on the $\varphi - X/CNT$ systems. We also mapped using the molecular viewer VESTA\cite{Momma:11}, the isosurfaces of the total charge density difference $\rho_{diff} (r)$, which is defined by Equation \ref{Deltarho}:

\begin{equation}
\rho_{diff} (r) = \rho_{SiW12@\varphi-X/CNT} (r)-\rho_{SiW12} (r)-\rho_{\varphi-X/CNT} (r)
\label{Deltarho}
\end{equation}

In this definition, $\rho_{SiW12@\varphi-X/CNT} (r)$ is the total charge density of the SiW12 cluster adsorbed at the functionalized or pristine CNT, $\rho_{SiW12} (r)$ represents the charge density of the SiW12 cluster; while $\rho_{\varphi-X/CNT}$ represents the charge density of the functionalized or pristine CNT. Each of the three components in Equation \ref{Deltarho} are considered to be at the same unit cell, separately.

\section{Results and discussion}
\label{Result}

\subsection{Strutural Description}
\label{sec:pbe_vdw}

The use of CNTs as electrodes has been used in energy storage devices, such as supercapacitors and Li-ion batteries. Specifically, at the experimental level, the presence of MWCNTs is to be expected due to their facile disposition at the bulk, instead of isolated CNTs. In this respect, a computational model of electrode with several concentrical CNTs would be of interest. Nevertheless, we consider that the implementation of a single-CNT would be sufficient to model the interaction with an external material. The latter is based upon ballistic transport and conductivity measured on MWCNTs \cite{Bandaru:07}, which appears to associate to MWCNTs a generalized metallic character. Consequently, it may be inferred that independently of the choice in the structural disposition of a single CNT (zig-zag or armchair), a metallic nature is to be expected in MWCNTs used as electrodes on energy storage devices. At the theoretical level, it results computationally more feasible to use a single CNT than two or more CNTs that may act as a MWCNT. In this respect, since a metallic character is expected in a MWCNT, the use of an armchair CNT was implemented in the present study. In this respect, we adopted a (7,7) armchair CNT in order to reproduce such metallic character. We chose those dimensions in the CNT to accommodate a single unit of POM that averages 1 nm of size. The interest to use SiW12$@$Carbon nanocomposites in energy storage devices are due to the fact that their redox processes are reveiled on more negative potentials compared for instance, to PMo12$@$Carbon materials. This behavior makes SiW12$@$Carbon nanocomposites, a potential electrode for its use as anode in energy storage devices, such as in batteries of asymmetric supercapacitors.

\begin{figure*}
\includegraphics[width=1.0\columnwidth,keepaspectratio=true]{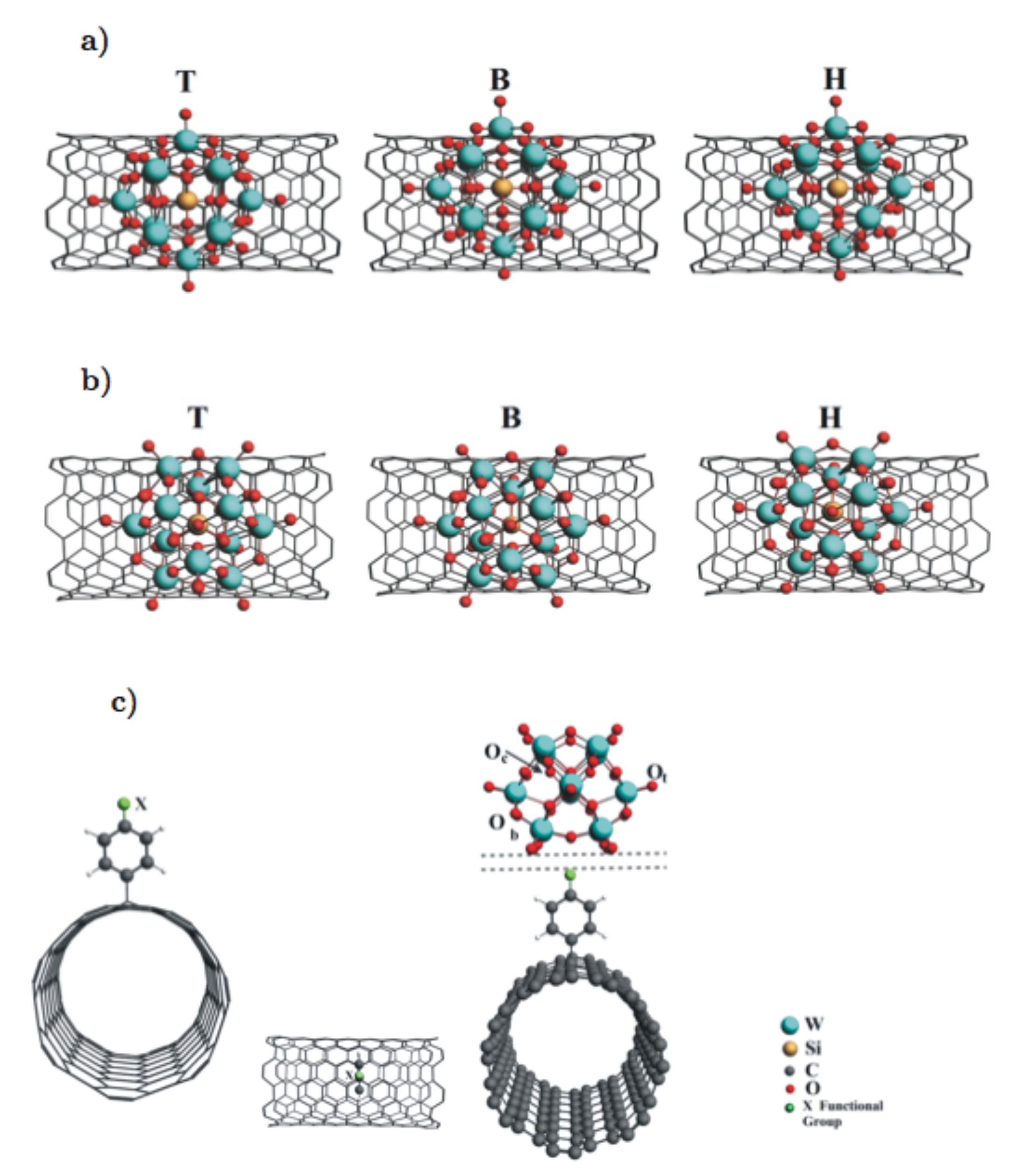}

 \caption{ \label{Models} Representation of models under study: SiW12$@$CNT top view. { \bf a)} S4 symmetry ;{ 
\bf b)} C3 symmetry; { \bf c)} $\varphi$-X$/$CNT and $SiW12@\varphi-X/CNT$, where $X$ denotes the functional groups under study: $-NH_2, -OH, -COH$ and $-COOH$. }
\end{figure*}

	We fully optimized the armchair model of CNT at DFT level using the functional and basis set previously described in the Computational Details. Considering the optimized structure, we deposited the POM SiW12 on the surface of the optimized CNT in a guess configuration located at 2.5 {\AA}, representing the approximated bond length at which POM systems and carbon matrices interact \cite{Wen:13,Wen:12}. As it has already been shown, two possible geometrical orientations are known, regarding how a Keggin-type POM may interact with a surface; namely, S4 and C3 symmetries\cite{Wen:13,Muniz:17}. Further, from each symmetry, 3 different orientations with respect to a graphene matrix have also been identified\cite{Wen:13,Muniz:17}, that is, 'Top', 'Bridge' and 'Hollow' orientations. Such orientations may be named T, B and H, respectively (see Fig. \ref{Models} {\bf (a)} and {\bf (b)} ). Such orientations refer to the central axis of the POM passing through a carbon atom (T-orientation), a pair of carbon atoms on a graphene layer (B-orientation) and at the center of a ring on the network (H-orientation), also on the graphene sheet.

It is important to denote that a composite material made of a POM cluster grafted on a pristine or functionalized CNT is expected to be neutralized due to the effect of a solvent\cite{Lang:17} when the system is $in$ $operando$ at energy storage devices. Such neutralization may be modeled by including explicit solvent effects in our calculations or by incorporating the presence of counterions to balance the total electronic charge\cite{Muniz:17,Lang:17} . We used a standardized methodology to model the presence of such counterions by introducing 4 Na$^{+}$ cations located close to the bridging oxygen atoms of the SiW12 POM (see {\bf Fig. S1} of SI). Consequently, all calculations (or stated otherwise) were performed with the presence of 4 Na$^{+}$ counterions. In order to explicitly account the Na$^{+}$ cations, we considered a basis set restricted to the $s$ orbital with a small cutoff radius of 0.497 Bohr. Besides, an equivalent real-space mesh cutoff of 250 Ry was introduced as suggested by Korstyrko et al\cite{Kostyrko:10} and Wen et al\cite{Wen:12a}. From this perspective, the Na atoms act in the calculations as point charges of unitary charge $+|e|$. The optimized structural parameters of the isolated SiW12 cluster are reported in {\bf Table S1} and compared with those parameters found in experiment \cite{Kong:06}. The theoretical parameters are in reasonable agreement with those given experimentally, the slight elongations ranging from 2\% to 6\% may be due to the solvent conditions where the SiW12 cluster parameters were experimentally measured \cite{Kong:06}.

Taking the latter into account, we performed 3 different relaxations at the C3 symmetry; that is, C3-T, C3-B and C3-H (see Fig. \ref{Models} {\bf (b)}). We also relaxed the composite SiW12$@$CNT system with S4-symmetry, and we found the geometries S4-T, S4-B and S4-H, as depicted on Fig. \ref{Models} {\bf (a)} . After computing the adsorption energies in accordance to Equation \ref{Eads}, it was found that the S4-H symmetry presents the lowest-energy configuration (see the results reported on Table \ref{Table_Ads_Ener}). This result is in close agreement with that found by Mu\~{n}iz et al\cite{Muniz:17} for the system SiW12$@$Graphene, where an S4-symmetry corresponds to the lowest-energy configuration. The magnitude of the interaction between the SiW12 cluster and the CNT at the S4-H symmetry is about -3.8 eV, which may be attributed to a non-covalent interaction of the electrostatic-type, and it is also consistent with the adsorption energy of -3.7 eV found for the SiW12$@$Graphene at S4-H symmetry. We found an averaged distance from the CNT surface to the POM of 187.1 pm (see Table \ref{Table_Interp_Dist}). 

\begin{table}
\begin{tabular}{|cc|}\hline
Model& Adsorption energy (eV)\\
\hline
SiW12@CNT & \\
S4-T & -3.66 (-5.13)\\

S4-B & -3.74 (-5.12)\\

S4-H & -3.83 (-5.10)\\

C3-T & -3.45 (-4.25)\\

C3-B & -3.29 (-3.98)\\

C3-H & -3.27 (-3.96)\\

\hline
SiW12@$\varphi$-X$/$CNT &\\
$\varphi$-NH$_2$ & -3.59 (-3.80)\\

$\varphi$-OH & -3.56 (-3.84)\\

$\varphi$-COH & -3.29 (-3.45)\\

$\varphi$-COOH & -21.33 (-21.64)\\
\hline
\end{tabular}
\caption{\label{adsorption} Adsorption energies of systems under study. $X$ denotes the functional groups $-NH_2, -OH, -COH$ and $-COOH$, respectively. The values in parenthesis correspond to those obtained using the Dion's VDW-DF functional \cite{Dion:04,Cooper:10}.}
\label{Table_Ads_Ener}
\end{table}

\begin{table}
\begin{tabular}{|cc|}\hline
Model& Bond length (pm)\\
\hline
SiW12@CNT & \\
S4-T & 190.0 \\

S4-B & 211.2 \\

S4-H & 187.1 \\

C3-T & 224.6\\

C3-B & 211.0 \\

C3-H & 236.4\\
\hline
SiW12@$\varphi$-X$/$CNT &\\
$\varphi$-NH$_2$ & 103.0 (111.6)\\

$\varphi$-OH & 111.5 (98.6)\\

$\varphi$-COH & 115.1 (121.5)\\

$\varphi$-COOH & 133.0 (143.0)\\
\hline
\end{tabular}
\caption{Interplanar averaged bond lengths of systems under study. $X$ denotes the functional groups $-NH_2, -OH, -COH$ and $-COOH$, respectively. The values in parenthesis correspond to those obtained using PBE+vdW functional \cite{Tkatchenko:09,Carrasco:14}. Note that those values with no parenthesis were computed at PBE level of theory.}. 
\label{Table_Interp_Dist}
\end{table}

We defined that distance as the average length among the external oxygen atoms $O_t$ (where $t$ holds for terminal) that belong to the SiW12 cluster and the CNT surface (such values are presented in Table \ref{Table_Interp_Dist}). The S4-H symmetry corresponds to the configuration where the POM cluster is adsorbed at the shortest distance. We also computed the adsorption energies at the VDW-DF level of theory \cite{Dion:04,Cooper:10} by considering the relaxed geometry previously computed at the PBE level. That is, we explicitly introduced dispersive effects of the vdW-type into our calculations. The results are presented in parenthesis in Table \ref{Table_Ads_Ener}. It may be seen that for the C3-symmetries, 17\% to 19\% of the adsorption energy is due to the vdW attraction contribution, while for the S4 symmetries, it ranges from 25\% to 29\%. In particular, the S4-H configuration is stabilized with a 25 \% of the adsorption energy due to this vdW interaction.

It was previously stated that the functionalization with organic groups onto the carbon substrates acting as electrodes, allows to enhance the density currents \cite{Cuentas:16} during the charge/discharge cycles observed in the voltammogram studies. This may be verified after the grafting of POMs at the composite CNT functionalized with organic groups. Furthermore, it has recently been shown \cite{Cuentas:16} that the $\varphi-NH_2$ and $\varphi-OH$ functional groups on Vulcan carbon with the presence of POMs, yields the same effect. Consequently, we performed a series of relaxations with the same CNT and a family of 4 organic functional groups, namely $i$) $\varphi-NH_2$; $ii$) $\varphi-OH$; $iii$) $\varphi-COH$ and $iv$) $\varphi-COOH$. The guess structures provided for the geometry optimizations were performed by considering the position of the organic functional groups at a top, bridge and hollow locations, taking into account the notation given above. The lowest-energy structure in all 4 cases is that located at a top-configuration, $i.e.$ above a carbon atom, which slightly deviates one of the carbon atoms (about 50 pm) from the wall's plane on the CNT, as depicted on Fig.\ref{Models} {\bf (c)}). We used this final disposition to graft a SiW12 cluster on the 4 functionalized cases of CNTs, previously stated.

Since the S4-orientation corresponds to the minimal energy configuration, we optimized all functionalized CNT geometries with the SiW12 cluster at the site of the functional group  (see Fig.\ref{Models} {\bf (c)}). The distances from the lowest plane formed by the terminal O$_t$ oxygen atoms and the functionalized CNTs are presented in Table \ref{Table_Interp_Dist} (see also Fig. \ref{Models} {\bf (c)} as a reference). It can be seen that all bond lengths range from 103.0  to 133.0 pm, representing shorter intermolecular distances than those found in the interaction of SiW12 cluster with the pristine CNT (see Table \ref{Table_Interp_Dist}). Despite the system $SiW12@\varphi-COOH/CNT$ reports an averaged intermolecular bond length of the same order of magnitude than those found for the other systems in the series, this is the only case with an enhanced attraction, comparable with a covalent bonding. That is, after computing the binding energies as given by Equation \ref{Eads}, it was evidenced that the attraction (see Table \ref{Table_Ads_Ener}) among the POM and the functionalized CNTs may be addressed to a non-covalent interaction of the electrostatic-type for systems: $SiW12@\varphi-NH_2/CNT$, $SiW12@\varphi-OH/CNT$ and $SiW12@\varphi-COH/CNT$ (see Fig. \ref{Models} {\bf (c)} ). The binding energies in these systems amounts to -3.6, -3.5 and -3.3 eV, respectively. Nevertheless, it was verified that for the system $SiW12@\varphi-COOH/CNT$, such binding energy is significantly strengthened to -21.3 eV, which may be remarkably considered as a bonding of the same order of magnitude of a covalent attraction. This is in agreement with the experimental insight found by Cuentas-Gallegos et al \cite{Cuentas:06}. They evidenced a large POM loading for CNTs functionalized with -$COOH$ group. In order to verify the contribution to the adsorption energy coming from a possible vdW attraction, we also evaluated Equation \ref{Eads} using the Dion's VDW-DF functional \cite{Dion:04,Cooper:10}. We assessed such values by considering the relaxed structure found at PBE level, and no geometry optimizations at VDW-DF level were performed to save computational time. Such values are also reported in Table \ref{Table_Ads_Ener} and given in parenthesis. Our calculations showed that the adsorption energies for the $SiW12@\varphi-X/CNT$ systems (with X = $-NH_2, -OH, -COH$ and $-COOH$) under study, are virtually unaltered with respect to the energies calculated at the PBE level. This certifies an electrostatic nature in the interactions. In particular, the system $SiW12@\varphi-COOH/CNT$ reports a value of -21.64 eV, which is virtually the same than that obtained at the PBE level. This indicates that the bonding for this particular case may not be considered as physisorption, and it directly evidences that the SiW12 cluster chemically interacts with the $\varphi-COOH/CNT$ susbstrate, giving rise to a covalent-type bonding. As it was previously stated \cite{Muniz:17}, the relaxations performed at the PBE + vdW level \cite{Tkatchenko:09,Carrasco:14}, provided us acceptable geometry optimizations of POM$@$Carbon substrate by considering no $Na$ counterions, since such scheme is not implemented at FHI-aims code. Nevertheless, the computation of the adsorption energies is overestimated for this reason, but without affecting the performance on the geometry optimizations \cite{Muniz:17}. In this regard, we fully optimized the 4 cases under study using the PBE+vdW scheme, and all relaxed geometries present virtually the same parameters than those computed with the PBE scheme using SIESTA code (those values are reported in parenthesis in Table \ref{Table_Interp_Dist}). That is, all averaged interplanar distances predicted within this methodology, present a difference ranging from 5\% to 11\% with respect to the values found with the presence of counterions. This indicates that the PBE functional is sufficient to correctly model the possible vdW interactions present in these systems. Consequently, we may assign to the $\varphi-COOH$ functional group, an enhanced attraction along the series of functional groups, and the composite system $SiW12@\varphi-COOH/CNT$ may represent a potential candidate to be implemented as an electrode material in energy storage devices, since the POM retention is a desirable condition to have an improved performance through the charge/discharge cycling.

\subsection{Electronic structure properties}

\begin{figure*}
     \centering

\includegraphics[width=1.2\columnwidth,keepaspectratio=true]{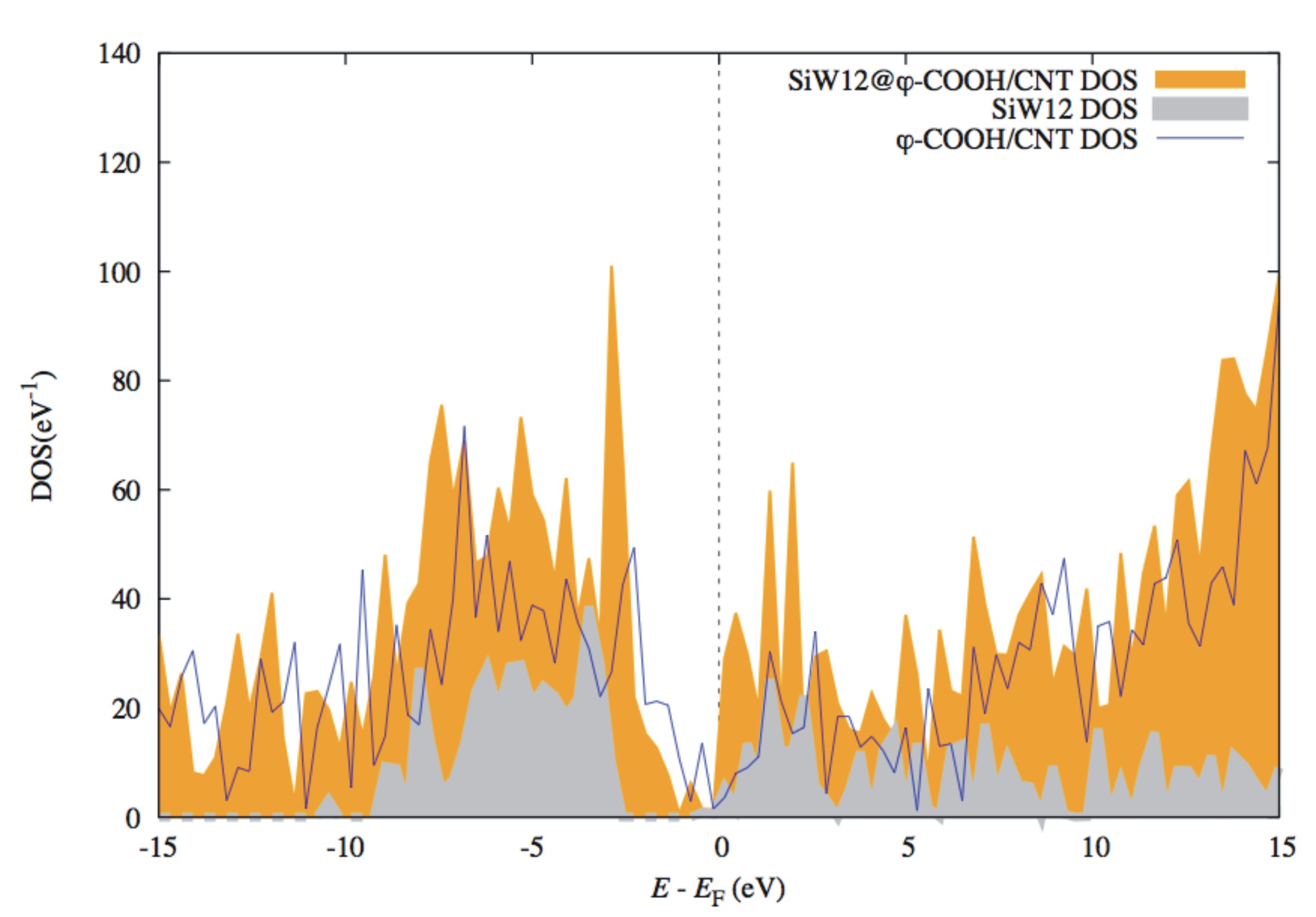}

     \caption{ \label{DOS} Density of states (DOS) and Projected Density of States for $SiW12@\varphi-COOH/CNT$
 system.}
\end{figure*}

All relaxed geometries in the family of composite materials were further characterized by considering their total Density of States (DOS) and Projected Density of States (PDOS). These results are presented in Fig. \ref{DOS} and

 {\bf Fig. S2-S13} of Supplementary Information (SI). {\bf Fig. S2} shows the DOS of the isolated (7,7) CNT, where the expected metallic character is evidenced due to the population of states at the Fermi level. On the other hand, {\bf Fig. S3} depicts the DOS and PDOS coming from the isolated SiW12 cluster. In this case, the main contribution to the total DOS is that due to the W atoms on the peripheral of the POM. Furthermore, we have also collected the DOS and PDOS of the functionalized CNTs. Such results are presented in {\bf Fig. S4-S7} of SI. It can be verified that the functional groups $i)$ $\varphi-NH_2$, $ii)$ $\varphi-OH$, $iii)$ $\varphi-COH$ and $iv)$ $\varphi-COOH$, provide new states to the total DOS and slightly reduces the metallic character of the CNT. Such states are those basically coming from the O atoms. This behavior is also altered if the SiW12 cluster is grafted on the pristine CNT (see {\bf Fig. S8} of SI), and the presence of the SiW12 triggers the rising of new states around the Fermi level. This may be produced due to the electrostatic attraction observed in this composite material (see Table \ref{Table_Ads_Ener}), which also reduces the metallic character of the CNT.

A general tendency in the series of composite systems $SiW12@\varphi-X/CNT$ is the modification on the total DOS around the Fermi level (see Fig. \ref{DOS} and {\bf Fig. S9 - S11} of SI). This may be indicative of the opennig of new pathways that the ions in the composite material may be able to track. Particularly, the rising of new DOS may be readily verified in case $iv)$ (as depicted in Fig. \ref{DOS}), that is, at the $SiW12@\varphi-COOH/CNT$ system. This may be attributed to one of the mechanisms yielding possible enhanced density currents in charge/discharge cycles performed on energy device processes, which may also be combined with its intrinsecal covalent interaction reported between SiW12 and $\varphi-COOH/CNT$. This gives composite system $iv)$, a unique signature among the other cases in the series.

In order to deepen into the understanding of the plausible mechanism for the POM retention and possible charge transfer at the interfase of SiW12 and $\varphi-X/CNT$, attributable to the different functional groups under study, we have calculated the projected density of states (PDOS) for the different $\varphi-X/CNT$ systems. {\bf Fig. S12} of SI shows the results obtained for PDOS due to the oxygen atoms at the functional groups $\varphi-OH$, $\varphi-COH$ and $\varphi-COOH$, attached to the CNT as well as those corresponding to the nitrogen atoms in $\varphi-NH_2/CNT$. It is also presented the contribution of carbon in the particular case corresponding to $\varphi-COOH/CNT$. Undoubtedly, the oxygen atoms correspond to the functional group involved and not to those at the SiW12 cluster, considering that these calculations were performed in the $\varphi-X/CNT$ structures and not in the $SiW12@\varphi-X/CNT$ systems. As can be seen from this graph, only the oxygen corresponding to the carboxyl group presents about 5 states/eV in the vicinity of the Fermi level. A peak of $\sim$ 6 states/eV corresponding to the carbon atom at the radical $-COOH$ in the $\varphi-COOH$ functional group is present at the Fermi level. Furthermore, the oxygen atoms of the $-OH$ and $-COH$ functional groups and the nitrogen atom of the amine group generate states far from the Fermi level located deep into the valence band and it would explain their weak contribution on the loading of the SiW12 cluster.

There is a difficulty, at least in appearance, to be able to identify with certainty the different contributions of the carbon atoms to the projected densities of states, since carbon is found as well in the carbon nanotube, as in the aromatic ring of the phenyl group and even in some functional groups $X$. Due to this argument, we have chosen to analyze the PDOS results due to hydrogen atoms, which are presented in {\bf Fig. S13} of SI. The hydrogen atoms are exclusively bounded to the carbon atoms present in the phenyl ring and in addition to the nitrogen in the particular case of the amine group. The smallest number of states/eV is explained by the lower number of valence electrons in the case of hydrogen. Again, only the $\varphi-COOH/CNT$ structure shows the states around the Fermi level populated. In the specific case $\varphi-NH_2/CNT$, a peak with $\sim$ 1 state/eV (located at approximately -4 eV) is attributable to the hydrogen atoms associated with the phenyl ring, and several more peaks (about five) are observed with almost null density of states. The delocalization of the electrons in the phenyl ring could be responsible for the latter. It is observed a density of states of about 4 states/eV located at $\sim$ -5.0 eV, which it is due to hydrogen atoms bonded to the nitrogen atom. Finally, we were able to confirm the hypothesis of the delocalization of the electrons in the phenyl ring, observing the presence of multiple peaks for the PDOS of the hydrogen atoms in the 4 cases under study. In this way, it can be confirmed that it is the functional group $X$ associated to the phenyl ring, and not the ring itself, that is responsible to modify the electronic structure properties of the systems under study.

\subsection{Charge transfer analysis}

\begin{figure*}

\includegraphics[width=1.2\columnwidth,keepaspectratio=true]{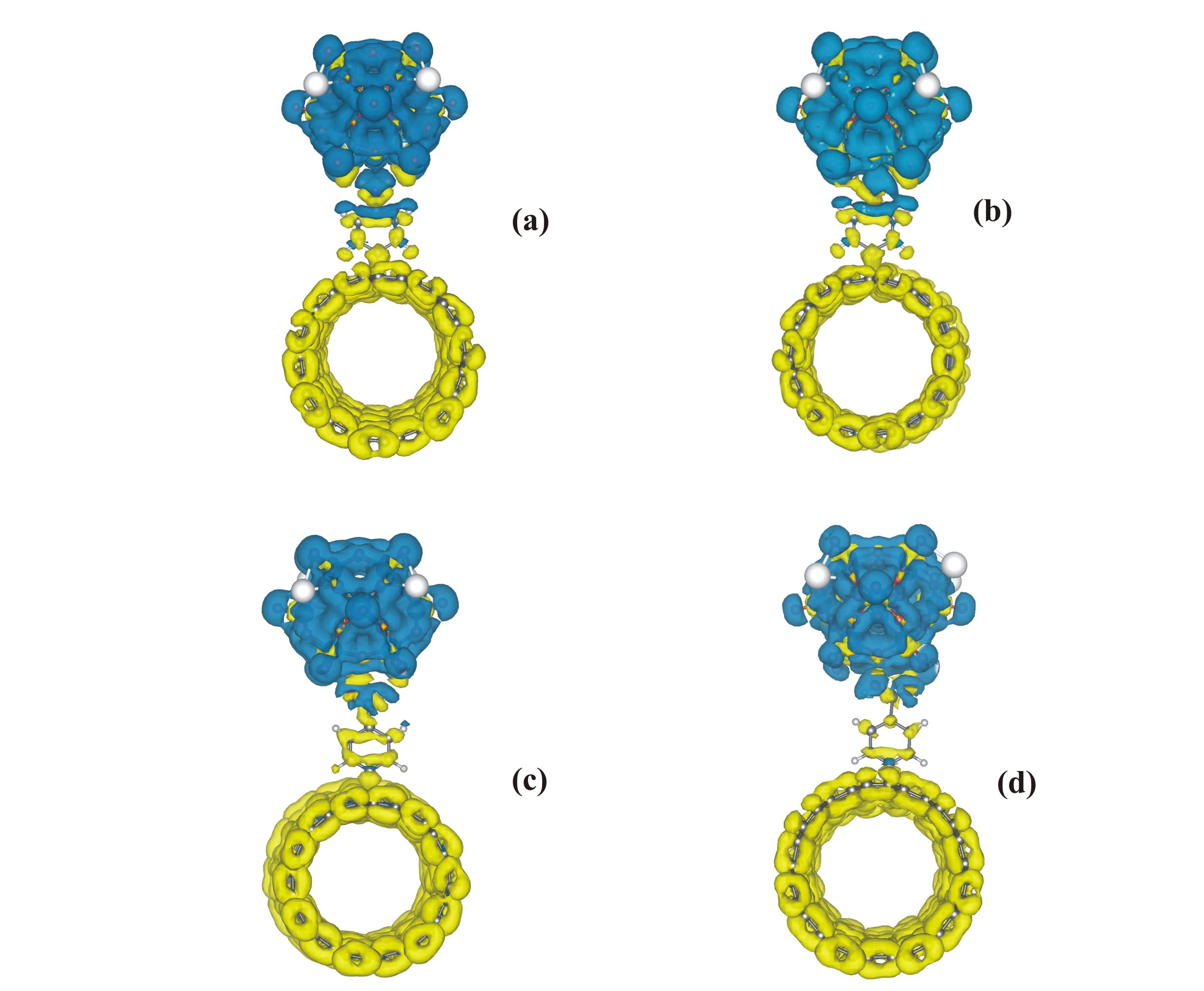}

\caption{\label{isosuperficies} Isodensity surfaces of charge density difference $\rho_{diff}$.  $\rho_{diff}$ 
for model  {\bf a)}  SiW$_{12}$@$\varphi$-NH$_{2} /$CNT, {\bf b)}  SiW$_{12}$@$\varphi$-OH$/$CNT,{ \bf c}  SiW$
_{12}$@$\varphi$-COH $/$CNT, {\bf d)}  SiW$_{12}$@$\varphi$-COOH$/$CNT.}

\end{figure*}

In order to give more insight into the nature of bonding among the SiW12 cluster and the functionalized CNT, we plotted the charge density difference isosurfaces $\rho_{diff}$ for the 4 cases under study in accordance to Equation \ref{Deltarho}, as given in the Computational Details section. The resulting isosurfaces are presented in Fig. \ref{isosuperficies} {\bf (a-d)}, the regions in yellow represent the locations where the electronic charge is depleted, and the regions in blue represent the locations where the charge was transferred to. In all cases, a clear charge

migration coming from the CNT to the SiW12 cluster was verified. This is a revealing result that may directly be associated to the SiW12 availability to retain electronic charge, which makes polyoxometalates unique chemical species. This is also in agreement with the results disclosed by Wen et al\cite{Wen:13} in the PMo12$@$CNTs systems. It is important to highlight that the charge transfer mechanism in the composite systems $SiW12@\varphi-X/CNT$ (where $X$ corresponds to the 4 functional groups) is opposite to that found in the SiW12$@$Graphene case\cite{Muniz:17}. In that work, our group evidenced that the charge transfer comes from the SiW12 system to the graphene sheet.

In the present work, the former behavior is also reported with the SiW12$@$CNT, $i.e.$ in the SiW12 system grafted on the pristine CNT, as it is depicted in {\bf Fig. S14} of SI. We presented the most stable configuration, corresponding to S4-H. In {\bf Fig. S14}, the yellow regions on the isosurface show the locations where the electronic charge was depleted and the blue regions show the sites where the charge is allocated. It is evident that the POM acts as an electronic sponge that may effectively retain charge after the cluster has been anchored to the substrate, and it is consistent with previous experimental works\cite{Liu:15}. This may suggest that this effect is present in a POM$@$CNT composite system, independently of the presence of a functional group.

The mechanism of charge transfer on SiW12$@$Graphene $vs$ SiW12$@$CNT may be ruled by the specific carbon geometry where POM interacts. That is, in the former instance, the planar extended graphene sheet allows the POM to freely deplete some of the excess of the electronic charge, while the constrained and bounded disposition given by the CNT restricts the charge transfer from the POM to the CNT, and the sponge-like behavior of the Keggin polyoxometalates prevails; $i.e.$, a significant charge retention mechanism domains the composite material and an imminent relocation of the electronic charge from the CNT to the SiW12 cluster is evidenced.

As it was stated above, the $\varphi-COOH/CNT$ system presents the strongest adsorption energy in the series of systems under study. This may directly be inferred from the excess of electronic charge transferred to the region between the cluster and the functionalized CNT (see blue regions on Fig. \ref{isosuperficies} {\bf (d)}). That is, the $\varphi-COOH$ functional group presents the largest contribution of transferred charge at the region of interaction between the cluster and the functionalized CNT. The latter may give rise to a more effective screening mechanism than in the other 3 functional groups that consequently induces a chemical interaction, comparable to the magnitude of a covalent bonding. The charge transfer mechanism was also assessed by numerical estimation of the charge transfer with the aid of the Mulliken population analysis. The charge differences ($\Delta Q$) between SiW12 and the CNT functionalized in the 4 cases are reported in Table \ref{Table_Charges}. The negative values in the table indicate a charge excess at the POM region that invariably takes place for all cases. These results were also benchmarked with other charge analysis schemes, such as those given by Hirshfeld\cite{Hirshfeld:77} and Voronoi\cite{Fonseca:03}. Independently of the chosen scheme, the same behavior for all cases was found, which certifies the charge depletion on the CNT as suggested by the charge density difference analysis.

\begin{table}[h]
  \begin{tabular}{|l|c|c|c|}
    \hline
    $X$ & 
    \multicolumn{3}{|c|}{ $\Delta Q$ ($SiW12@\varphi-X/CNT$)} \\
    \hline
     & Mulliken & Hirshfeld & Voronoi\\
     \hline
     $-NH_2$ &  -1.00 & -0.99 & -1.06 \\
     \hline
     $-OH$ &  -0.95 & -0.95 & -0.99 \\
     \hline
     $-COH$ &  -0.95 & -0.95 & -0.99 \\
     \hline
     $-COOH $ &  -0.92 & -0.99 & -0.99 \\
     \hline
     $pristine $ &  -1.28 & -1.26 & -1.31 \\
     \hline
  \end{tabular}
  \caption[Table caption text]{Charge differences located at the SiW12 cluster adsorbed on a CNT with $X$ = $-NH_2, -OH, -COH$ and $-COOH$ or $X$ = No functional group (pristine case).}
  \label{Table_Charges}
\end{table}

\subsection{Related-energy storage capabilities on the series of systems under study}

In order to shed light into the energy storage properties that may be further addressed in the $in$ $silico$ design of energy storage devices, we performed a series of calculations in accordance to Equation \ref{EadsH}:

\begin{equation}
E^{H}_{X-ads} = E_{X-H}-E_{X}-E_{H},
\label{EadsH}
\end{equation}

\begin{figure*}
\centering
     \includegraphics[width=1.2\columnwidth,keepaspectratio=true]{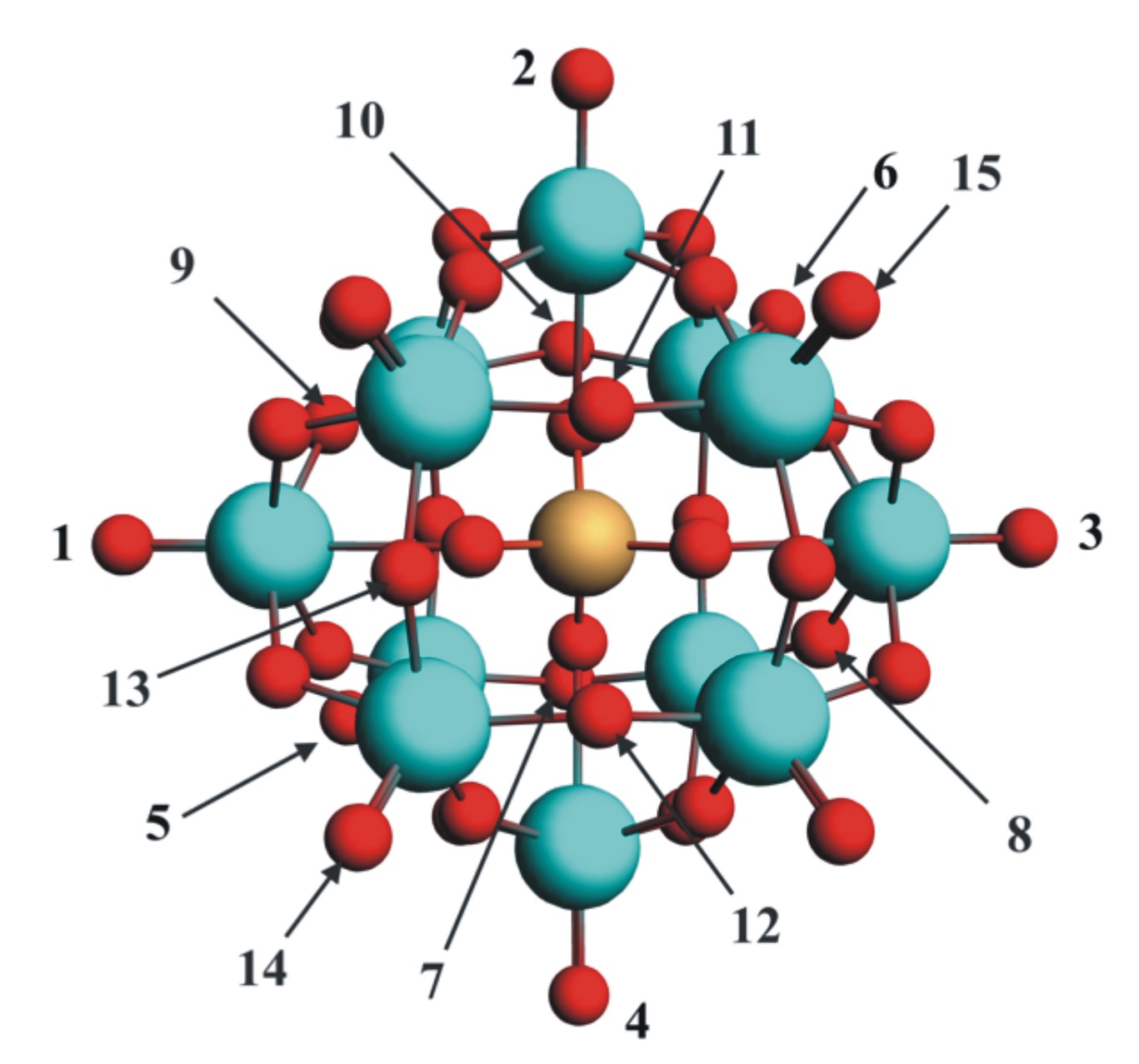}
     \caption{ \label{Locations_adsortion_hidrogen} Test points where the test charges were located to evaluate
 the adsorption energies in the systems $SiW12@\varphi-X/CNT$, where X = $-NH_2, -OH, -COH$, $-COOH$, and the pristine case.}
\end{figure*}

In this equation, $E^{H}_{X-ads}$ is the adsorption energy of an adsorbed test charge (a Hydrogen atom) at a given location, $E_{X-H}$ is the total energy of the composite system with the adsorbed H atom; $E_X$ is the total electronic energy of an $X$ composite system in the absence of the test charge and $E_H$ is the total energy of an isolated H-atom in the same unit cell used in all calculations. The $X$ notation in Equation \ref{EadsH} considers the possibility to introduce the SiW12 grafted on the pristine CNT or on the cases where it is grafted at the functionalized CNTs. We applied the methodology given by  Mu\~{n}iz et al\cite{Muniz:16}, where a test charge (provided by a H-atom) is located on the peripheral of the POM along its equatorial axis at the configurational geometry of equilibrium, and it may also be extended to other locations around the POM, as it is shown in Fig. \ref{Locations_adsortion_hidrogen}, which was formerly computed with the provided methodology. The allocation of such test charges may be arbitrarily distributed around the neighborhood of the POM. Such charges were accomodated as depicted on Fig. \ref{Locations_adsortion_hidrogen}, since we are interested to qualitatively describe the charge retention mechanism in the presence of the 4 different functional groups under study.

\begin{figure*}
\centering
     \includegraphics[width=1.2\columnwidth,keepaspectratio=true]{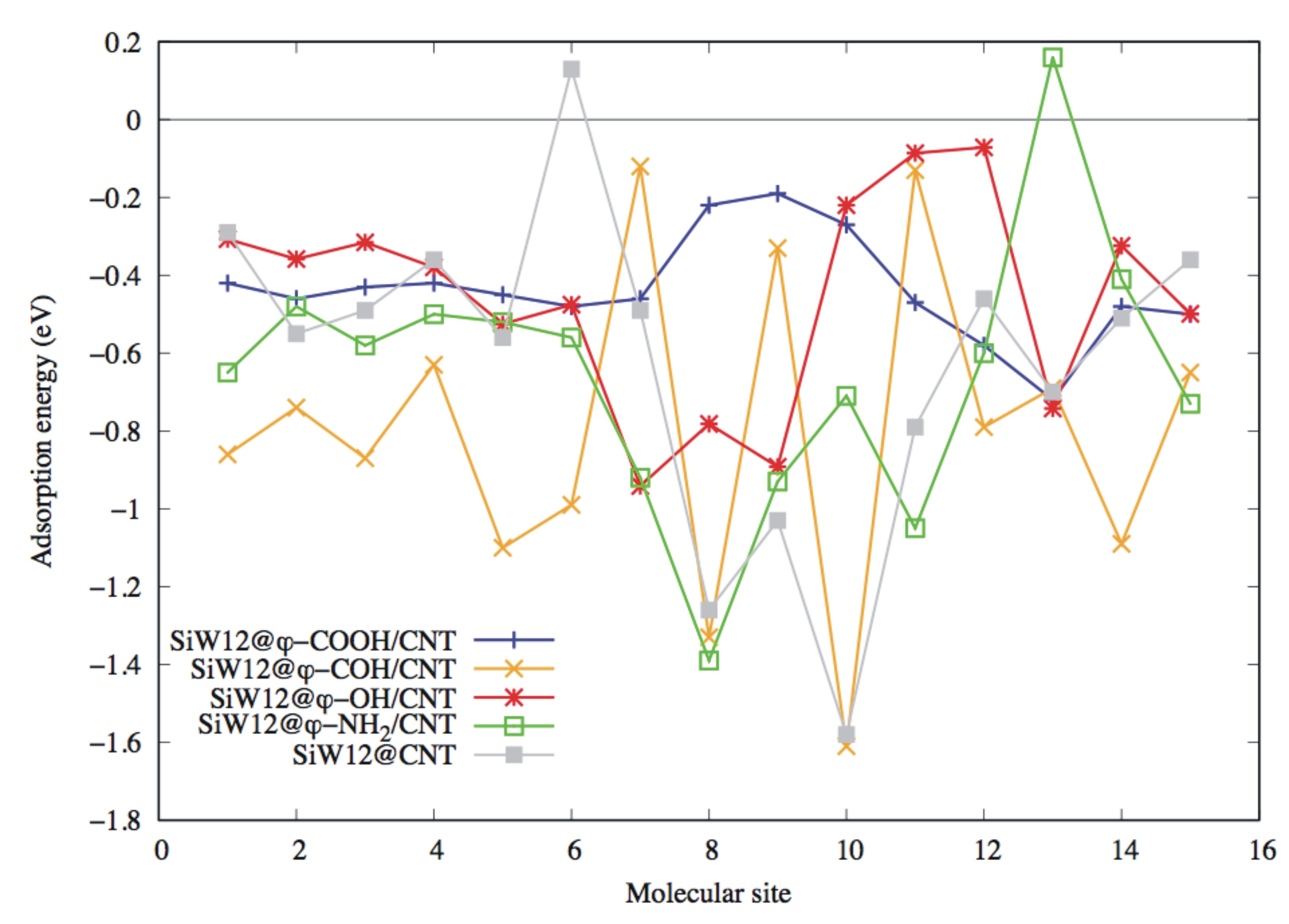}
     \caption{ \label{adsortion_hidrogen} Adsorption energies of test charges at the different sites allocated 
around systems $SiW12@\varphi-X/CNT$, where X = $-NH_2, -OH, -COH$, $-COOH$, and the pristine case.}
\end{figure*}

The comparative results are presented in Fig.\ref{adsortion_hidrogen}. It may be seen that the SiW12$@$CNT system (S4-H symmetry) with no functional groups anchored to the carbon surface, reports adsorption energies (as given by Equation \ref{EadsH}) randomly distributed, which may induce a weak charge retention after the SiW12 cluster is anchored to the CNT surface. On the other hand, all 4 cases with the different functional groups are also presented in Fig. \ref{adsortion_hidrogen}. Besides, all abrupt fluctuations found in the functionalized systems $(i)$ to $(iii)$, may indicate that a charge retention mechanism would be less effective than that found for sytem $(iv)$, namely $SiW12@\varphi-COOH/CNT$, which in average, remarkably maintains stable adsorption energies throughout all test sites. This indicates that this composite system may be a potential candidate to be implemented as an electrode in energy storage devices such as supercapacitors or Li-ion batteries due to its capacity to retain electronic charge on the region surrounding the real space of the SiW12 cluster. Furthermore, the remaining functional groups; namely $\varphi-NH_2$, $\varphi-OH$ and $\varphi-COH$, may also be used for the same applications, since the interaction energies around the selected locations are attractive as those found for the $\varphi-COOH$ functional group. Nevertheless, this particular system, may be considered as the most promising material for this purpose.

\section{Conclusions}

A systematic theoretical study was performed on a series of organic functional groups grafted on the surface of a CNT. The anchoring of the SiW12 polyoxometalate was also studied in order to assess the contributions of bonding coming from covalent and dispersive-type interactions. It was found that the $SiW12@\varphi-COOH/CNT$ system rules the interactions with an attraction of the same order of magnitude than that found in a covalent bonding, among the other systems in the series. The carboxylic functional group dominates the bonding through the family due to its unique availability to retain electronic charge, which strongly anchors the SiW12 to the functionalized CNT. The bonding of SiW12 on the rest of functional groups $\varphi-NH_2$, $\varphi-OH$ and $\varphi-COH$, is governed by non-covalent attractions of the electrostatic-type with a contribution coming from a van der Waals interaction. The charge transfer mechanism in these composite materials appears to be ruled by the geometry of the carbon substrate, since it was evidenced that the electronic charge is transferred from the CNT and it is allocated at the SiW12 cluster, while at planar geometries such mechanism is reversed. The evaluation of charge retention reveals that the presence of an organic functional group increases the possibility to adsorb electronic charge around the SiW12 cluster in the composite material, and makes the system $SiW12@\varphi-COOH/CNT$, a potential candidate material to be implemented as an electrode in energy storage devices. The methodology presented in this work may aid in the search of novel functional groups that allow POM retention on carbon substrates more efficiently.

\label{Concl}

\section*{Acknowledgements}
\label{Acknow}

 A.G.L. wants to acknowledge the PhD. Scholarship provided by CONACYT with No.306891. N.E.T. is grateful for the Posdoctoral Scholarship provided by CONACYT with Project No.229741. J.M. wants to acknowledge the support given by C\'atedras-CONACYT (Consejo Nacional de Ciencia y Tecnolog\'ia) under Project No. 1191; the computational infrastructure provided by Laboratorio Nacional de Conversi\'on y Almacenamiento de Energ\'ia (CONACYT), and the Supercomputing Department of Universidad Nacional Aut\'onoma de M\'exico for the computing resources under Project No. LANCAD-UNAM-DGTIC-310. The authors acknowledge the computing time provided by Universidad Polit\'ecnica de Chiapas at the Laboratorio Multidisciplinario de C\'omputo de Alto Rendimiento. The authors would like to acknowledge the financial support given by DGAPA (Direcci\'on General de Asuntos del Personal Acad\'emico) under Project No. IN112414. The authors thank helpful discussions with Dr. Marina Rinc\'on and Dr. Marcelo Lozada.  

 \section*{References}
 \label{Ref}


\bibliographystyle{elec_comm_MS} 
\bibliography{gold_gen}







\pagebreak










\end{document}